\documentclass[amsmath,aps,10pt,amssymb,rmp,floatfix]{revtex4-1}
\usepackage[english]{babel}
\usepackage[colorinlistoftodos]{todonotes}
\usepackage{multirow}
\usepackage{comment}
\usepackage{relsize}
\usepackage{float}
\usepackage{textcomp}
\usepackage{silence}
\usepackage{xcolor}
\usepackage{graphicx}
\usepackage{epstopdf}
\usepackage{color}
\usepackage{dcolumn}
\usepackage{bm}
\usepackage[mathlines]{lineno}
\usepackage{threeparttable} 
\usepackage{booktabs} 
\usepackage{siunitx}

\setcitestyle{round}
\begin{document}
\textbf{ \Large{ \textit{Journal of UAP Studies} \quad\quad\quad\quad\quad\quad\quad  Society for UAP Studies}}
\begin{center}
\Large{\textbf{\underline{Proceeding for the SUAPS 2024 Conference (Phys.~Sci.~Workshop)}}}
\end{center}
\begin{figure}[H]
\centering
\includegraphics[width=0.52\textwidth]{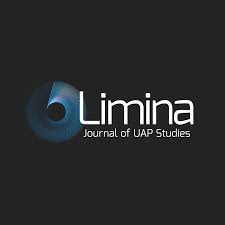}
\end{figure}

\title{How much time do we have before catastrophic disclosure occurs?}  
\date{\today}
\author{Matthew Szydagis}
\affiliation{Dept.~of Physics, UAlbany SUNY, \\ 1400 Washington Avenue, \\
Albany, NY 12222-0100 USA \\
Member, SCU, the Scientific Coalition for UAP Studies \\
Member, UAPx}

\begin{abstract}
\noindent 
Claims of the retrieval of crashed craft or vehicles from non-human intelligence(s) (NHI) abound in the popular culture and media. For this article, the number of unsubstantiated claims is utilized to estimate the time expected until a ``catastrophic disclosure'' occurs. The term was defined at the 2023 Sol Foundation’s inaugural conference as an accidental disclosure of strong evidence of the existence of NHI. The phrase refers to this occurring outside the control of major human institutions, such as governments and militaries.~One possible example of this is the crash of a piloted (space)craft or ET probe in the middle of a busy metropolis (such as the city square, \textit{e.g.}, New York City's Times Square).~The distribution of humans across the Earth’s surface, the population as a function of time, and the fraction of individuals owning a camera-phone, also versus time, are each taken into consideration as a foundation for a rigorous statistical analysis. This author adopts a skeptical and agnostic approach and does not conclude NHI or ET are real, but applies standard statistical distributions as educational examples of critical thinking to an issue which captures the imagination of the public as almost no other issue does. Making the extraordinary assumptions that sentient species other than humans exist, are capable of constructing vehicles for transportation, and are sufficiently fallible that their technology can malfunction, it becomes possible to quantify some potential answers to the question of how long it might be before smartphone imagery and/or video evidence appears on the web and becomes irrevocable via classification in the modern era. Results of simulations of numerous potential scenarios with varying degrees of optimism indicate that, if NHI are real, catastrophic disclosure may actually happen accidentally rather soon, with the mean expected year being 2040 $\pm$ 20 under the default assumptions.
\textbf{}
\end{abstract}

\maketitle

\section{Introduction, and Literature Review}
\vspace{-5pt}
The concept of the existence of conscious, sentient, and intelligent races of non-humans has captivated the collective imagination of humanity for centuries, or longer~\cite{Thigpen2022}. The issue of their exact nature(s), and the question of technological capability enabling visits to the Earth, are separable from the question of existence. UFOs (Unidentified Flying Objects) and UAP (Unidentified Aerospace or Anomalous Phenomenon/a) are often conflated with the notion of spacecraft that are being piloted by NHI (non-human intelligence) including and especially ET (extra-terrestrials), in spite of these terms, especially the latter newer one, referring only to an unknown phenomenon, or phenomena, which may include anomalous atmospheric effects which are naturally occurring but are simply not yet understood~\cite{UAPxJPAS}. Such conflation happens for a good reason, however: the measured kinematics, specifically high velocities and accelerations, of at least some small fraction of observed, but non-identifiable, aircraft~\cite{Knuth+etal:2019}. Since late 2017 especially, discussions about UFOs, as well as about aliens, have once again been thrust into the limelight, within the mainstream media~\cite{NYT:2017}. 2023--2024 ``whistleblower'' claims served to reignite discussions. New and serious instrumented studies have appeared~\cite{Cloete,Szenher_2023,Watters2023}.

The ET hypothesis is arguably rational. Exploration of it can easily be justified by discoveries of many~thousands of exoplanets, probably constituting only a small sample from billions or even trillions, with $O$(10\%) potentially habitable according to an anthropocentric habilitity criterion. That is based upon a host-star separation permitting the presence of liquid water, and oxygen in the atmosphere, thus not even counting life not-as-we-know-it, nor exo-moons~\cite{KASTING1993108,GuillermoGonzalez,Gallet2017}. A relatively recent example of discovery of multiple possibly~human-habitable worlds, announced~via a NASA press conference, is the TRAPPIST-1 planetary system~\cite{Gillon2017}. They are a sample from an estimated 300 million~\cite{Bryson_2021}. While the relativistic time dilation and length contraction are known to work in favor of high-speed travelers, the problems of fuel and of propellant with sufficient thrust for long-term high acceleration, shielding against cosmic radiation and (fatally) Doppler blue-shifted starlight, and celestial navigation remain unresolved, at least by contemporary human beings. A civilization comparable in age to the galaxy might become capable of overcoming all these engineering (not physical) difficulties involved, but that is purely optimistic speculation, with no complete explanation for a ``hide and seek'' type of behavior, so it is best to adopt a model-independent (non-Drake) approach to the question of NHI origin(s) and travel abilities.

As a result, this paper will make no assumption in that regard -- NHI, if assumed to be real and able to reach Earth, may be traditional aliens, come from other dimensions or universes, or be intelligent lifeforms that co-evolved next to humans and are thus also native to Earth. Furthermore, we do not discount human explanations, \textit{e.g.}~time-travelers or present-day breakaway civilizations, nor the mundane explanations of many sightings. All of the quite wild hypotheses with no solid empirical evidence as yet for any of them are covered very well in other sources, such as \cite{ITSanderson} and \cite{HPuthoff}. The only (extraordinary) assumptions being made for this research article are: 
\vspace{-2pt}
\begin{enumerate}
    \item NHI (or, a very advanced yet unknown group of humans possessing extraordinary vehicles) actually exist.
    \vspace{-5pt}
    \item Regardless of point of origin or motivations: they possess high-tech craft operating on or near Earth's surface.
    \vspace{-5pt}
    \item They possess some degree of fallibility, making accidents such as unplanned (\textit{i.e.}, crash) landings realistic.
\end{enumerate}
\vspace{-2pt}
\noindent
Given the 3 simple points above we can now ask the question of when good preliminary evidence would be captured by ordinary civilians, who have reported many thousands of strange sightings in the sky~\cite{Antonio}. Our focus, however, will be on the catching of crashes with smartphone camera technology, through random chance, considering an additional, fourth assumption of an annual crash rate that can be grounded by a review of the literature regarding UFO crash claims. Though we failed to find any scholarly papers from (external and blind) peer-reviewed, high-impact researched-focused journals in the mainstream scientific community for this particular sub-topic (nor many on UFOs in general due to the enduring stigmas~\cite{GStahlman}), there is the initial effort of \cite{Maristela}. But the most useful resources for alleged crashes were: \cite{Schmitt1991,Randle1994,Randle1995,Randle2010}.

At the Sol Foundation symposium organized by Prof.~Garry Nolan of Stanford and held on November 17--18, 2023, some (intelligence-community) speakers used the phrase ``catastrophic disclosure'' for the scenario where the military-industrial complex is not the player driving disclosure, but scientists, engineers, citizen-scientist researchers, and even ordinary citizens (with the unspoken postulate being that human governments know a great deal more than they are disclosing to date). Many of the Sol speakers claimed that in such a scenario the impacts upon our society (on politics, religion, etc.) of disclosure of the existence of NHI would be more catastrophic if compared to a slow, controlled, and planned version of it. This article will present skeptical but not debunking analyses. It should allow readers, including relevant politicians and lawmakers, to estimate when catastrophic disclosure could transpire on its own. On the other hand, it can serve to disprove the most extreme claims of crash rates, especially as time goes by without disclosure transpiring. The statistical methods employed here may be germane to setting upper limits on the rate of occurrence of many different kinds of ``exotic'' phenomena. The particular case of a publicly-confirmed crashed NHI craft would most likely constitute the single most important discovery in the history of science, if not all human history in general.

\section{Methods}
\begin{figure}[ht]
    \centering
    \includegraphics[width=0.975\textwidth]{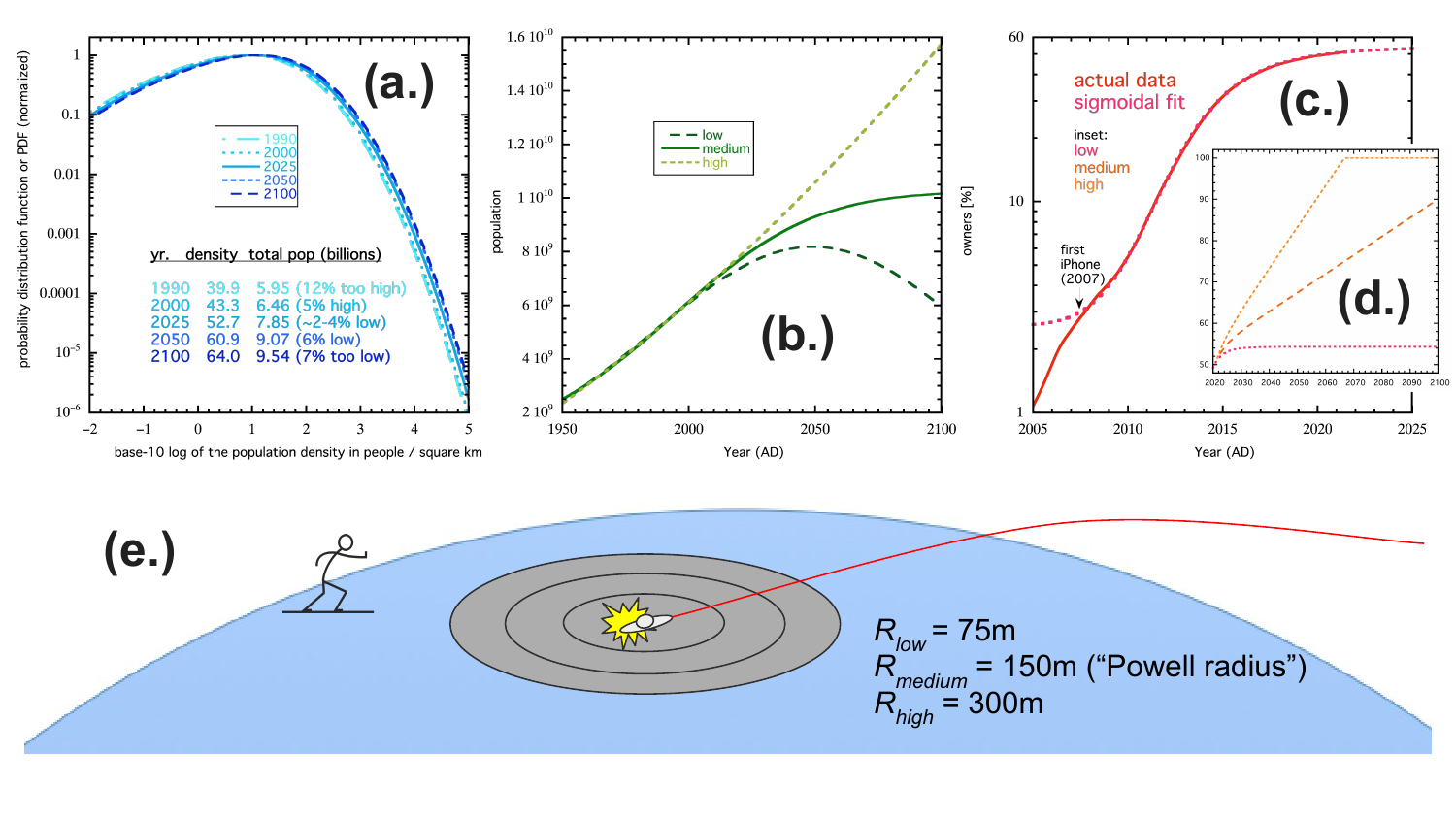}
    \vspace{-24pt}
    \caption{A summary of all of the numerical inputs to the results. (a.) An amplitude-normalized PDF for the population density at several example years. (When the PDF is normalized by area instead, its peak is at 1.56 not 1.00 on the y-axis.) Both plot axes are logarithmic. Because the x-axis is cut-off at -2, \textit{i.e.}~0.01 people/km$^2$, there is a slight overestimate of total populations  in the earliest years, although this is irrelevant since they are too early. A slight underestimate in later years was produced by cutting x off at +5, or 100,000 people/km$^2$, which exceeds Macao (in China) and Monaco's present-day values, of about 20,000 people/km$^2$ (higher densities may be reasonable for the future, even if rare). (b.) The 3 United Nations projections applied to the non-normalized versions of the curves in the left graph, which uses only ``medium'' as an example, to set the proper totals. These correspond to the 3 formulae in Eqn.~(2), a symmetric Gaussian (which unphysically collapses to 0 pop, but not for many centuries), an asymmetric sigmoid which asymptotes to just over 10 billion, and a quadratic polynomial, for the low, medium, and high cases, respectively. (c.) Actual data on smartphone ownership percentage as a solid red line~\cite{SmartPhone50}, with a symmetric-sigmoid curve fit to the log of the fraction in pink dots. When the data are plotted using a log-y scale, an inflection point which may be attributable to the introduction of the Apple iPhone is evident. Our sigmoidal fit unphysically asymptotes to 2.6\% at left, but the fit is only applied to the post-2007 time period. (d.) The inset displays 3 possible futures: a value barely above 50\% persists (pink dotted line again) and (linear) increases to 100\% at 2 different slopes -- a dark orange long-dashed line and light orange short-dashed line, the latter unrealistically reaching 100\% in the late 2060s, but this is simply one, extremely optimistic, scenario, but justified based on publicly-claimed slopes~\cite{SmartPhone,SmartPhoneHigh} (e.) Differing radii used as low, medium, high scenarios to define minimum distances within which crashes are obvious~and~safely~capturable. No thought is given to matters like object size or camera focal length, but a factor-2 uncertainty on radius should cover most such issues.}
    \label{fig:1}
\end{figure}

\vspace{-5pt}
The mathematical formulae applied in the analyses presented herein, while findable in the code, are also summarized here, and demonstrated in Figure~\ref{fig:1}. They bear a marked similarity to those applied to camera captures of low-density wildlife~\cite{RareWildLife}. First, the world's population density profile probability density was empirically fit~using:
\vspace{-11pt}
\begin{align}
    e^{\frac{-(x-\xi)^{2}}{2 \times 1.711^{2}}} [ 1 + \mathrm{erf} ( -2.4 \frac{x-\xi}{1.711 \sqrt{2}} ) ];~\mathrm{x~is~log_{10}(density).}~\xi=1.5467+ 4.2773\cdot10^{-11} P;~\mathrm{P~is~overall~pop.} \\
    P = 8.183\cdot10^9 e^{\frac{-(t-2049.1)^{2}}{2 \times 63.746^{2}}}\mathrm{,}~1.027\cdot10^{10} + \frac{2.96\cdot10^8-1.027\cdot10^{10}}{[ 1+(\frac{t}{2020.9})^{56.388}]^{1.988}}\mathrm{,}~4.0181\cdot10^{11} - 4.7813\cdot10^8 t + 1.4014\cdot10^5 t^2,
\end{align}
\vspace{-3pt}
\noindent
with three $P$-increase scenarios indicated from left to right (low, moderate, and high) and $t$ the time (in years). The base-10 log of the camera smartphone ownership fraction (dimensionless) vs.~$t$ was modeled as an asymptotic S-curve:
\vspace{-6pt}
\begin{align}
-0.26506 + \frac{-1.5883+0.26506}{1+[\frac{(t-1999.5)}{12.387}]^{6.7706}}~\mathrm{where,~\textit{e.g.},~-0.26506~means~10^{-0.26506}=54.3\%~phone~ownership.}
\end{align}

\noindent
A radius $R$ defined a circle (approximated as flat) on the surface of the Earth, within which at least one individual with a phone is located, based on a Poisson-varied number of people, based upon a mean density drawn from a skew-Gauss distribution (Equation~1). The default used was 0.150~km, or the Powell Radius, because of the crude~approximation mentioned in \cite{PowellUFOs} of 500~ft.~for reliable eyewitness testimony (sans images), but for the sky, not the ground.

Not only the planetary population as a function of time, projected into the future based on UN projections~\cite{Raftery2014}, but also the distribution of persons across the approximately 1.49 $\times$ 10$^8$ km$^2$ of land~\cite{PopDensityHist}, were taken into consideration. An average person density is not used to represent the entire globe. This would be unrealistic for areas such as Antarctica at one extreme, and large, dense cities on the other (with New York city not even being the densest). No preference for visitation location was considered. The probability density function (PDF) for population density in units of people per km$^2$ was modeled as a skew-Gaussian function, skewed in favor of lower densities, and spanning $10^{-2}$--$10^{+5}$ individuals/km$^2$. The PDF peaks at $\sim$10/km$^2$, with an average of $\approx$30--60/km$^2$, depending on the year being modeled. A robust density profile was required for this study, but the only scholarly one found was for 1998~\cite{PopDensityHist}. It was adjusted for later years by smoothly varying just the centroid of the skew-normal distribution to approximate new density distributions with time. This approximation, with all areas effectively going up in person density uniformly, was validated by integrating under the resulting curves and verifying that one recovers the correct total populations. Future work should account for the width and skew of the probability density changing over time as well, although adjusting these would be overkill for a zeroth-order analysis. (Note that, for reproducibility, all exact equations and numbers can be found in the downloadable C code.)

The fraction of people who own smartphones (with cameras) versus time was also necessary to model; however,~for simplicity we ignored any bias toward greater ownership in higher-population-density areas, applying only a flat~value. That said, a Poissonian distribution, the most common assumption in STEM for the modeling of rare events, was used to simulate local variation in density of phone owners, with a Poisson random number generator likewise implemented to simulate the number of events (on land) each year (2008 and later), with rate estimates discussed later used only to set the Poisson means. Even though the Poisson function can be well approximated using a Gaussian or ``normal'' distribution (bell curve) at high rates, it has the advantages over that of producing only non-negative, integer values, and, because its mean and variance are equal, a separate value for width is unneeded, unlike with normal or log-normal functions, with the latter recommended by \cite{maccone2022evosetimathematicaltoolcladistics} for simulating similar problems, but with a free-parameter variance. While the first ``smartphone'' was invented during the early 1990s (IBM), and the first camera-phones during the late 90s or in 2000, no year earlier than 2008 was considered in our analyses, a year after introduction of Apple's iPhone, which led to a greater explosion in ownership, with competing companies also making phones with cameras.

To run Monte-Carlo simulations of random, smartphone-driven disclosure, three levels of close-by crash rates were taken as benchmarks constituting simplistic, order-of-magnitude, Fermi-problem-style estimates -- 1, 10, and 100 per century. The lowest value essentially comes from treating only the Roswell incident~\cite{Schmitt1991,Randle1994,Corso} as a (potentially) real example of an NHI spacecraft crash, the sole one for the entire twentieth century, as suggested by K.~Randle, who has stated that most other incidents were probably hoaxes and misidentifications. Our middle-of-the-road value of 10 stems from taking the claims of whistleblower David Grusch at face value~\cite{Grusch}. The extremum of 100/century or 1/year originates from \cite{Randle2010}, which contains a list of 118, with most not NHI-related however, as already stated, thus making 100 not just the highest rate assumed within this work, but likely also the least realistic. That being said, \cite{Wood2024} cites over 50 possibilities, using a rating scheme to judge the probabilities of their veracity, and Randle has written that lists of well over 300 alleged crashes exist. Therefore, 2 is a reasonable power of 10 for the upper end of our Fermi estimation.

\vspace{-11pt}
Other authors may consider different distributions like log-normal, but we start with Poisson here. The probability~of 1 crash can be Taylor-expanded as $r$, but probability $p$\footnote{The very same form as Eqn.~(4) can model the $k$ people who see and successfully record an incident, replacing $r$ with the mean number of people $N$=$10^{x} \pi R^2$ multiplied by the camera-phone ownership fraction of Eqn.~(3). ($x$ is the log density, and (people/km$^2$)$\times$km$^2$=people)} of $k$ crashes in a given year is fully, correctly expressed as:

\begin{equation}
p = \frac{\langle r \rangle ^k}{k!} e^{-\langle r \rangle},~\mathrm{where~\langle r \rangle~is~the ~expectation~value~for~the~annual~crash~rate,~and~k=0~or~a~positive~integer.}
\end{equation}

\noindent
For instance, for the rate of 10/century or $\langle r \rangle$ = 0.1/year, the $p$ of $k$=0 is highest, at 0.90 or 90\%, with a 9.0\% chance of $k$=1, and 0.45\% for $k$=2, and so forth. The virtual dice can be re-rolled once a day, week, month, or year in code, by adjusting the units on $r$, with no substantive change in the ultimate results. We opted for the annual timeframe for computational speed, and performed $10^5$ trials for each of our 18 (3*3*2) cases based on 3 possibility combinations (low, medium, high) from this section combined with 3 crash rates, and 2 different starting years, 2008 and 2024, for initializing the simulations. The large number of trials (10$^5$) ensured that statistical uncertainties would be negligible for our final results, dominated instead by the systematic uncertainties of our choices of the quantitative assumptions. A 2008 start was used to validate our work, checking the probabilities that disclosure should already have happened.

\section{Results}
\begin{figure}[hb]
    \centering
    \includegraphics[width=0.92\textwidth]{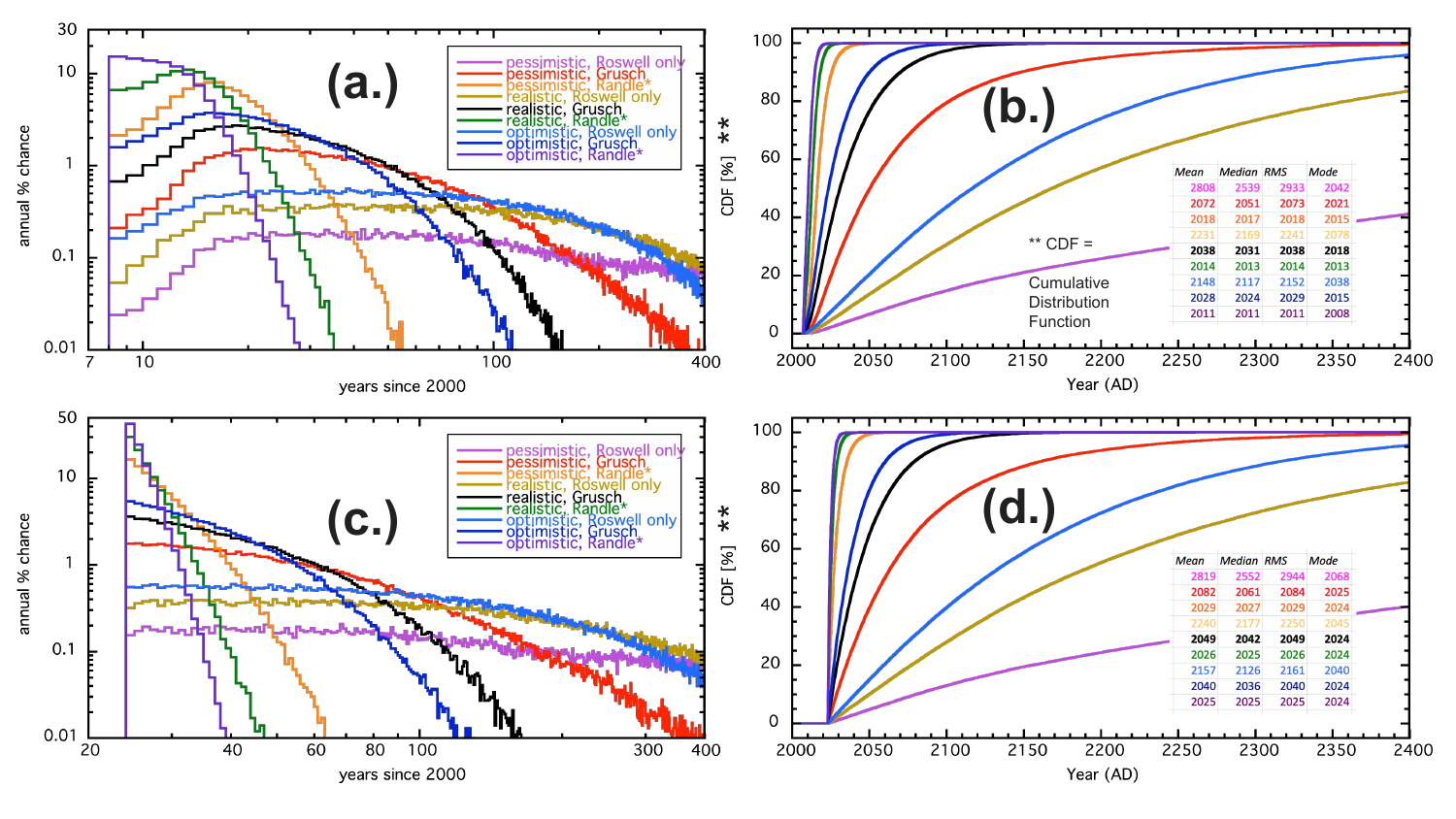}
    \vspace{-15pt}
    \caption{Odds of catastrophic disclosure by year at left (a,c) on log axes, and summed at right (b,d), beginning in 2008 (a,b), and 2024 (c,d). Pessimistic refers to all inputs ``low'' (Fig.~\ref{fig:1}), realistic medium, optimistic high. Roswell means 0.01 crash/year, Grusch 0.1/yr, and Randle 1, but * indicates Randle thinks that is far too high. Some statistics are tabulated at the right.} 
    \label{fig:2}
\end{figure}

\vspace{-5pt}
The simulated year of an accidental disclosure of NHI existing is surprisingly soon for many combinations of inputs, or predicted to be a year from the past even, despite the decades-old prevalence of non-smartphone cameras, private satellite companies, and many other signal channels available to civilians all not having been taken into account: see Fig.~\ref{fig:2}. For pre-2024 results, we can use the strange outcome to rule out certain cases. 2011.4 $\pm$ 2.8 (mean plus/minus standard-deviation uncertainty $\sigma$) for violet implies that 1 crash/year paired with $R_{low}$ is ruled out at a level of nearly 5$\sigma$. The green case (1 crash a year and $R_{med}$, the Powell Radius default) of 2013.7 $\pm$ 3.9 is nearly 3$\sigma$ discrepant with our reality of non-disclosure. The expectation value for orange (1 crash/year and $R_{high}$) of 2017.9 $\pm$ 6.4 differs from 2024 by only $\approx $1$\sigma$, so there is no tension with reality there. Remarkably, the majority of the tested cases resulted in a 50\% cumulative probability of catastrophic disclosure by 2050 AD. No distinction was made regarding air/space-craft with visually obvious ``biologics'' (Grusch) at UFO crash sites, versus the crashes of automated probes (or, some thing ``in between'' for which our human categories, like organic and living versus robotic, AI, and artificial, are inadequate).

We are, however, assuming for all of our simulation results that captured evidence is qualitatively conclusive, without attempting to quantify what is meant by ``conclusive'' evidence in terms of video and/or photographic quality, duration or number of pictures, etc. We also postulate that one eyewitness with a phone is sufficient, due to the fact that they can rapidly share data with friends and family via e-mail and by text attachments, and on public-facing video-sharing websites such as YouTube, TikTok, or Vimeo, before the military or other authorities can rush to the scene to remove all physical evidence, then classify the incident. (Such actions have naturally been alleged by UFO conspiracy theorists going back decades.) Furthermore, having one initial witness should be sufficient due to one's ability to call to others in high-population-density areas and text friends within areas of any density (this criterion can be easily modified in the code, however, and the effect on the result explored). While humans have always been social creatures, there was no texting nor camera-phones in pockets in Roswell, NM in 1947, nor in Kecksburg, PA in 1965, for instance.

Figure~\ref{fig:2} presents all the sim results, divided by annual probability and cumulative (integrated) probability vs.~time, and also separated by start year, 2008 or 2024, with cases having the latter as the input essentially taking the current state of affairs (lack of proof on the web) as a (Bayesian) prior. In those latter situations, over-optimistic postulates can artificially generate enormous immediate peaks in probability, for 2024--2025. Using our ``central'' postulates, the predictions are 2038 $\pm$ 24 (2008 sim start) and 2049 $\pm$ 23 (2024 start) for the year of the initial but ``incontrovertible'' evidence being shared on the internet, assuming survival of strict checks of AI fakery. The errors (uncertainties) quoted here are non-Gaussian -- they are simply raw standard deviations $\sigma$ of asymmetric data. The mean expectations are not the same as the median, RMS (root-mean-square) or mode (peak in probability), with that last parameter having no value higher than 2068 even when others extend into the 22$^{\mathrm{nd}}$ century, due $e.g.$~to a global-population peak.
\vspace{-11pt}

\section{Discussion and Conclusion}
\vspace{-10pt}
If NHI are real then the correct question to ask is not IF disclosure can be forced, but WHEN. Figure~\ref{fig:2} right shows the probability always asymptoting to 100\% eventually -- when inputting the defaults, the chances (black curves) are 14--42\% by 2027 and 39--59\% by 2036. (Two values are quoted for each example year because of the differing beginning years.) This is because of the remarkable interconnectedness of the modern world, with cellular phones in individuals' pockets and bags nearly everywhere, even in developing nations, even when one makes the conservative claim of only a fixed 54\% maximum for the percentage of human beings utilizing camera-enabled smartphones. We~have~not~accounted for people who have multiple phones and subscriptions, but this could be the cause of the significant tensions between the sources we cited earlier ($\sim$50 as opposed to about 75\%). Because of the contradictions, it is important that distinct scenarios were studied. A further refinement to our study which readers can implement on their own, as the~code~has been provided with the paper, is the mixture of low, medium, and high inputs in place of an artificial correlation. It was done here for simplicity. Future work building on this paper by this author and/or others could also account for daylight versus night-time hours, terrain type, attention spans, and other such factors~\cite{MedinaSciRepKirk}, but these analyses presented here are a first stab in the literature as far as the author knows, in terms of a journal publication.

An unspoken assumption baked into every analysis within this manuscript is that we are referring to unmistakable, clear photos of crashes on land only, potentially less ephemeral than UAP in the sky, including if not especially over water~\cite{ITSanderson,PDennett}. The land accounts for only 29.2\% of the surface area of Earth. Therefore,~$\langle r \rangle$ = 0.1/year (our middle-benchmark crash rate) becomes $>$ 0.3/year for the entire globe, with $>$ 0.2/year for only the water-covered regions. If one equates all NHI with aliens and takes the estimated time frame of the first discovery of Earth by one space-faring race capable of interstellar travel as $\sim$1 million years~ago~\cite{Knuth2025},~then this implies over 200,000 defunct (extraterrestrial) vessels sitting on the bottom of oceans, lakes, and rivers~--~if~there was no change in technology over time. This is a surprising number motivating further searches like that described in \cite{loebOcean}, and this could even be an underestimate: one species may inform others of Earth and humanity's presence and progress, possibly leading to a time-dependent $r$-value (increasing?) driven by others' scientific curiosity.

Although this work may read as if it is a warning to government officials with solid knowledge of some NHI (again-- should they really exist) that their time grows short to maintain full control of the narrative, it should not just sound alarm bells. The mathematics is neutral, in the sense that one could also capitalize on the procedures contained here to debunk NHI and UAP crash retrieval claims, as more years without  disclosure pass by. The statistical techniques employed in this article can and should also be applied to cryptids, orbs, Earth lights, ball lightning, and any sort of ``paranormal'' claim that presently can boast of no proof that has convinced the majority of the scientific community, nor the public at large. Using the dark matter as an example, a mainstream scientific topic with an enormous number of indirect \textit{i.e.}~observational clues coming from cosmology and astrophysics, but no conclusive, direct evidence in the laboratory as yet, we note one can set limits on the probability of the interaction of new particles with normal matter composed of atoms~\cite{Aalbers_2023}. A scientist can set a limit on the rate of occurrence of any rare phenomenon with similar techniques, but without necessarily ruling it out completely. Not all exotic claims are false, as our history has repeatedly shown (atoms, germs, continental drift, meteors, air/space flight, nuclear power/bombs, relativity,...)

That being said, it would be a mistake to state that all initially ``crazy'' ideas have eventually been proven correct in the history of science, as such a claim would be far from accurate. However, subjective experiences do not constitute unambiguous final evidence in the physical sciences, making the raw data sets of well-calibrated scientific instruments absolutely critical to possess, not just information from the ``human sensor,'' witnesses with potentially faulty memories (adequate for legal systems and useful in the humanities and the social sciences, but insufficient in a physical science).

A central argument herein is that a phone may generate adequate evidence of an anomaly. Yet it is not a scientific instrument, so that is a major weakness we recognize in our own argument. While evidence from a single camera may convince the general public, it is not likely to convince most academics who continue to be unmoved by the existent plethora of UAP imagery. Having a video instead of photo(s) of a nearby crash on the ground, with a witness moving around and achieving different viewing angles, minimizes the possibility of an AI-generated hoax as a solution at the very least, especially if signs with letters and human fingers are also seen in the background~\cite{AIBadatWords}.

Moreover, a smartphone camera can collect what Prof.~Garry Nolan referred to at the Sol Foundation 2023 inaugural meeting as ``pre-data,'' a step above anecdotal non-data at least, which can still be used for corroborating scientific data, and justify choices of sensors. But, without funding and publications in mainstream journals, progress will still be difficult. A ``smoking gun'' phone video could precipitate increases in both of those, but would probably~not~be a substitute for the study of crash parts in person, to look for evidence of NHI technology, such as advanced unknown alloys, or isotopic concentrations inconsistent with our solar system~\cite{NOLAN2022100788}. The latter could be observed using mass spectrometry, or non-destructively with NAA (neutron activation analysis)~\cite{NAA}. An initially agnostic approach to data-taking is good~\cite{UAPxJPAS} but scientists must be allowed, sans fear of reputation loss, to entertain exotic hypotheses (like NHI), considering if data favor them or not~\cite{VillarroelCrash}.

\vspace{-10pt}
\section*{Acknowledgments}
\vspace{-5pt}
The author acknowledges Prof.~Eric Davis as well as David Grusch for inspiring this paper, and Mick West for his skepticism, and xkcd 1235. He thanks Prof.~Kevin Knuth for first getting him interested in UAPs, and Gary Voorhis, leader of UAPx. He thanks Profs.~Brenda Denzler, Beatriz Villarroel, and Doug Buettner for key criticisms of drafts.
\vspace{-25pt}
\bibliographystyle{chicago}
\bibliography{uap}
\vspace{-13pt}
\appendix
\section{Supporting Code}
\vspace{-13pt}
The C++ file catDisc.cpp is part of a paper download for the benefit of the technically-minded reader with knowledge of computer programming. It can, for instance, be easily compiled on Unix with the terminal command ``g++ -Ofast catDisc.cpp -o catDisc.out'' assuming one has the g++ compiler installed (the -Ofast optimization flag is optional).
\vspace{-13pt}
\section{Supporting Plots}
\vspace{-13pt}
This appendix contains an additional figure, to explore a more fine-grained variation of the crash rate, as opposed to just looking at three orders of magnitude, and of $R$.

\begin{figure}[ht]
    \centering
    \includegraphics[width=0.94\textwidth]{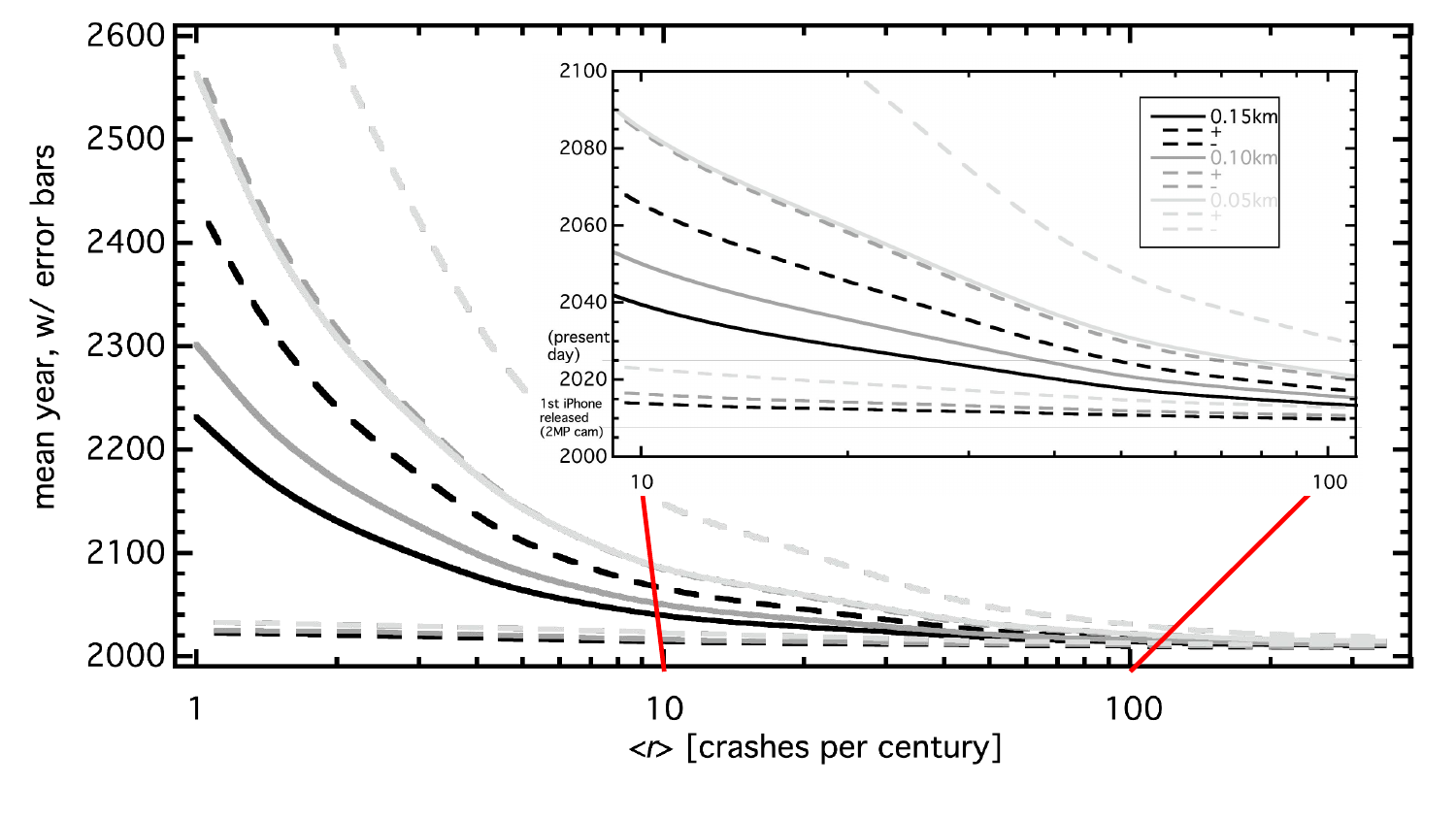}
    \vspace{-24pt}
    \caption{The year of accidental disclosure with uncertainty as a function of $\langle r \rangle$, with all other numerical assumptions held~fixed, at their ``medium'' levels: for human population growth, smartphone penetration into the population, and visibility radius (in black: $R$ is varied in gray). The inset is a zoom-in for 10--100 crashes/century. $r>20$ is disfavored by sims at the Powell~Radius (0.15 km). Discrete sim results are well fit continuously using y = 2008 + $m_1/r^{m_2}+m_3/r^{m_4}$, with both powers floating point. Smaller radii allow for the possibility of physical evidence retrieval by the crash witnesses, and may make hoaxing much harder, but the assumption of a single eyewitness was maintained: it is changeable to any number, however, in the code that is provided.} 
    \label{fig:3}
\end{figure}

\end{document}